# Room temperature spin-orbit torque efficiency and magnetization switching in SrRuO$_3$-based heterostructures

Sheng Li,[1,2,3,*] Bin Lao,[1,2,*] Zengxing Lu,[1,2,*] Xuan Zheng,[1,2,4] Kenan Zhao[1,2], Liguang Gong[1,2], Tao Tang[1,2], Keyi Wu[1,2], Run-Wei Li[1,2,3,†] and Zhiming Wang[1,2,3,‡]

[1]CAS Key Laboratory of Magnetic Materials and Devices, Ningbo Institute of Materials Technology and Engineering, Chinese Academy of Sciences, Ningbo 315201, China
[2]Zhejiang Province Key Laboratory of Magnetic Materials and Application Technology, Ningbo Institute of Materials Technology and Engineering, Chinese Academy of Sciences, Ningbo 315201, China
[3]Center of Materials Science and Optoelectronics Engineering, University of Chinese Academy of Sciences, Beijing 100049, China
[4]New Materials Institute, University of Nottingham Ningbo China, Ningbo 315100, China

[*]These authors contributed equally to this work
[†] runweili@nimte.ac.cn
[‡] zhiming.wang@nimte.ac.cn

**Abstract**

Spin-orbit torques (SOTs) from transition metal oxides (TMOs) in conjunction with magnetic materials have recently attracted tremendous attention for realizing high-efficient spintronic devices. SrRuO$_3$ is a promising candidate among TMOs due to its large and tunable SOT-efficiency as well as high conductivity and chemical stability. However, a further study for benchmarking the SOT-efficiency and realizing SOT-driven magnetization switching in SrRuO$_3$ is still highly desired so far. Here, we systematically study the SOT properties of high-quality SrRuO$_3$ thin film heterostructuring with different magnetic alloys of both IMA and PMA configuration by the harmonic Hall voltage technique. Our results indicate that SrRuO$_3$ possesses pronounced SOT-efficiency of about 0.2 at room temperature regardless of the magnetic alloys, which is comparable to typical heavy metals (HMs). Furthermore, we achieve SOT-driven magnetization switching with a low threshold current density of $3.8 \times 10^{10}$ A/m$^2$, demonstrating the promising potential of SrRuO$_3$ for practical devices. By making a comprehensive comparison with HMs, our work unambiguously benchmarks the SOT properties and concludes the advantages of SrRuO$_3$, which may bring more diverse choices for SOT applications by utilizing hybrid-oxide/metal and all-oxide systems.

## Introduction

Current-induced spin-orbit torques (SOTs) provide an efficient way to manipulate magnetization states for the potential magnetic memory and oscillator applications. SOTs originate from charge-spin conversion with strong spin-orbit coupling (SOC), which is intensively studied in heavy-metals (HMs) over a decade due to their potentially useful SOT-efficiency and low resistivity [1-4]. In addition to HMs, recently, a wide range of material systems are explored for realizing efficient SOT devices including topological materials [5-8], 2D transition metal dichalcogenide [9-11], and transition metal oxides (TMOs) [12-23]. In particular, TMOs with perovskite structure attract tremendous attention due to their rich electromagnetic properties and high SOT-efficiency closely related to their exotic electronic structures [24-28]. Due to the intimate entanglement among multiple degrees of freedom, including charge, spin, orbit, lattice, the electronic structure and the associated SOT-efficiency can be precisely engineered in high quality TMO films at atomic scale [29,30]. Therefore, TMOs provide a wide platform for searching and designing materials in future energy efficient and scalable multifunctional spintronics. [31-33].

Among oxide materials, 4d $SrRuO_3$ (SRO) is an outstanding member of perovskite family due to its low resistivity at ambient temperature, as well as high chemical and thermal stability [34,35]. These properties have been exploited as the electrodes for complex oxide heterostructures, such as Josephson junction, magnetic tunnel junctions, and capacitors [35-37]. Currently, it is recognized that SRO holds large Berry curvature and Weyl nodes, leading to lots of novel electromagnetic phenomena such as chiral anomaly induced negative magnetoresistance and anomalous Hall effect [28,38-40]. This intrinsic Berry phase of SRO, moreover, has been demonstrated to provide strong and high tunable SOT properties [14-16]. The spin Hall conductivity as large as $9 \times 10^4$ $\hbar$/2e S·m$^{-1}$ [15], and can be further manipulated via strain-controlled crystalline structure engineering. However, the reported spin Hall conductivity associated SOT-efficiency $\xi_{SOT}$ sometimes exhibit inconsistent magnitudes. For example, the $\xi_{SOT}$ can differ by several times in the same SRO/Py heterostructures grown on $SrTiO_3$ substrates [16]. Thus, it's desirable to benchmark the SOT-efficiency of SRO through a systematical study. Moreover, as a crucial step for realizing SOT-driven device, current-induced magnetization switching has not been demonstrated in SRO so far.

In this work, we systematically investigate the SOTs properties at room temperature based on SRO/magnetic-alloy heterostructures with different magnetic easy axes of in-plane and perpendicular to plane (IMA, PMA). To benchmark the SRO system, SOT-efficiencies and spin Hall conductivity of SRO are separately evaluated in both IMA and PMA configurations by harmonic Hall voltage technique. Besides, we demonstrate the SOT-induced magnetization switching in the PMA system, and the threshold current density could be one order of magnitude smaller than those in HMs. Finally, we compare the evaluated SOT properties with that of the previous reported SRO and HMs, and highlight the advantages of SRO as a spin source in oxide-metal hybrid and all-oxide systems for various SOT applications.

## Results

### $SrRuO_3$/Py and $SrRuO_3$/CoPt hybrid structures

High-quality SRO thin film with a thickness of 20 nm is coherently grown on (001)-orientated $SrTiO_3$ (STO) substrate by pulsed laser deposition (PLD, see details in Method Section). Structural properties of the film are determined by high-resolution X-ray diffraction (XRD) measurements. Figure 1(a) shows the $\theta$-$2\theta$ scan of the SRO film, in which distinct (001) and (002) peaks of both film and substrate implying the epitaxial growth of the film. The well-defined Kiessig fringes indicate a sharp SRO/STO interface and flat surface. According to the peak position, the out-of-plane lattice constant is calculated as 3.942 Å by Bragg's Law, which is larger than the bulk value 3.925 Å, indicating an in-plane compressive strain in the deposited film. Figure 1(b) exhibits the epitaxial relationship between the SRO film and substrate characterized via X-ray reciprocal space mapping (RSM) along (103) direction, in which the identical $Q_X$ values (marked by a red dash line) strongly suggest that the film is fully strained by the substrate. The atomic force microscope (AFM) result (Figure 1(c)) demonstrates an atomically flat surface with clear terraces, which is in consistent with the XRD results. After preparation of the SOT-source SRO layers, two magnetic components, i.e. 7-nm permalloy ($Ni_{0.81}Fe_{0.19}$) with pure IMA and amorphous 3.5-nm CoPt ((Pt(1)/Co(0.5)/Pt(0.5)/Co(0.5)/Pt(1))) with pure PMA, are individually deposited on the bare SRO by PLD and sputtering, respectively. Thereafter, two different SOT-switching systems with IMA and PMA, i.e. SRO/Py and SRO/CoPt, are prepared, as shown in Figure 1(d) and (e) respectively.

**SOTs-efficiency of IMA system**

The harmonic Hall voltage measurements are performed to characterize the SOT associated properties in the SRO/Py sample with IMA [11,41,42]. Figure 2(a) schematically presents the sample structure, directions of applied charge current $I$ and external magnetic field $H$ in our measurement configuration. The first and second harmonic Hall voltages, $V_{xy}^{1\omega}$ and $V_{xy}^{2\omega}$, are acquired simultaneously while rotating the angle $\varphi$ between $H$ and $I$ [43-45]. Figure 2(b) shows typical $V_{xy}^{1\omega}$ signals as a function of $\varphi$. Since the $V_{xy}^{1\omega}$ is equivalent to conventional Hall voltage, the obtained $\cos 2\varphi$ dependence indicates that the magnetization in the IMA sample is always aligned in plane during the measurements. The $V_{xy}^{2\omega}$ provides information about the spin current induced SOT, which cause small precession of the magnetization $M$ about its equilibrium position against external field. Correspondingly, the dampinglike and fieldlike components can be quantitatively analyzed via the $V_{xy}^{1\omega}$ and $V_{xy}^{2\omega}$ results by following formulas:

$$V_{xy}^{1\omega} = V_{PHE} \sin 2\varphi, \tag{1}$$

$$V_{xy}^{2\omega} = -\frac{1}{2} V_{DL} \cos \varphi + V_{FL} \cos \varphi \cos 2\varphi, \tag{2}$$

$$V_{DL} = \frac{V_{AHE}}{H+H_K} H_{DL}, \tag{3}$$

$$V_{FL} = \frac{V_{PHE}}{H}(H_{FL} + H_{Oe}). \tag{4}$$

Here, $V_{PHE}$, and $V_{AHE}$ represent the planar Hall voltage and anomalous Hall voltage. $H_{DL}$ and $H_{FL}$ denote the effective field of dampinglike and fieldlike components, which correspond to $\cos\varphi$ and $\cos\varphi\cos 2\varphi$ dependences. $H_K$ is effective anisotropy field of Py estimated to be 6500 Oe by the anomalous Hall measurement. $H_{Oe}$ denotes the Oersted field which has the same symmetry with $H_{FL}$. By fitting the measured $V_{xy}^{2\omega}$ data to Eq. (2), we obtained amplitudes of the two components varied with $\varphi$. Typical curves under $I$ = 3.5 mA and $H$ = 1000 Oe are shown in Fig. 2(c). Accordingly, as shown in Figs. 2(d) and 2(e), the extracted dampinglike and fieldlike voltages, $V_{DL}$ and $V_{HL}$, exhibit linear dependence against $1/(H+H_K)$ and $1/H$ for $I$ ranging from 1.5 to 3.5 mA. This indicates that our data can be well explained by Eqs. (3)-(4), thereby the $H_{DL}$ and $H_{FL}$ are reasonably estimated from the slopes, $V_{AHE}H_{DL}$ and $V_{PHE}(H_{FL}+H_{Oe})$, respectively. Particularly, $H_{FL}$ is obtained by subtracting the contribution of $H_{Oe}$, where $H_{Oe} = \mu_0 J t_{SRO}/2$ is derived from Biot-Savart Law. Here $J$ and $t_{SRO}$ are current density and thickness of the SrRuO$_3$ layer. Figure 2(f) shows the estimated effective fields as a relationship of $J$, where $H_{DL}/J$ and $H_{FL}/J$ are calculated to be 13.81±0.28 Oe/(10$^{11}$ A/m$^2$) and 12.32±0.05 Oe/(10$^{11}$ A/m$^2$). Finally, we evaluate the dampinglike (fieldlike) SOT-efficiency and spin Hall conductance via $\xi_{DL(FL)} = (2e\mu_0 M_s t_{FM}/\hbar)(H_{DL(FL)}/J)$ [19,46] and $\sigma_{SH} = \xi_{DL} \cdot \sigma_{SRO}$, where $M_S$ = 597 emu/cm$^3$ and $t_{FM}$ denote saturation magnetization and thickness of the magnetic layer, $\sigma_{SRO}$ = 64.1×10$^4$ S·m$^{-1}$ is conductance of the SRO layer. The values of $\xi_{DL}^{IMA}$ and $\sigma_{SH}^{IMA}$ are 0.175 and 11.2×10$^4$ $\hbar$/2e S·m$^{-1}$, comparable to the recent studies of SRO and the representative heavy metals, such as Pt [47-49], W [50], and Ta [51,52]. The $\xi_{FL}^{IMA}$ is estimated to be -0.003, which is negligible in this system.

**SOT-efficiency of PMA system**

To systematic confirm the SOT associated properties of SRO, the harmonic Hall measurements are alternatively carried out in the SRO/CoPt sample with PMA [53,54]. The measurement geometry is schematically depicted in Fig. 3(a), where the AC current $I$ is applied along $x$ direction, magnetic easy axis of the CoPt aligns toward $z$ (perpendicular) direction. During the measurements, the magnetic field $H$ is applied along either $x$ (longitudinal) or $y$ (transverse) direction to estimate the effective fields $H_{DL}$ and $H_{FL}$ by acquiring the first and second harmonic voltages $V_{xy}^{1\omega}$ and $V_{xy}^{2\omega}$. Because in the PMA configuration, the $H_{DL}(H_{FL})$ is parallel (perpendicular) to the $I$ in the $xy$ plane, and can be expressed as follows:

$$H_{DL(FL)} = -2 \frac{B_{DL(FL)} \pm 2\eta B_{DL(FL)}}{1-4\eta^2}, \tag{5}$$

$$B_{DL(FL)} = \left\{\frac{\partial V_{xy}^{2\omega}}{\partial H} \bigg/ \frac{\partial^2 V_{xy}^{1\omega}}{\partial H^2}\right\}_{H_{DL(FL)}}, \tag{6}$$

where $\eta$ = 0.105 is the ratio of $R_{PHE}$ to $R_{AHE}$ used to exclude the contribution from the planar Hall effect. We firstly confirm the AHE loop of the SRO/CoPt sample, as shown in Fig. 3(b), which exhibits the typical characteristic of PMA with a sharp coercivity field of about 100 Oe. To estimate the $H_{DL}$, $V_{xy}^{1\omega}$ and $V_{xy}^{2\omega}$ are measured simultaneously by sweeping $H$ along $x$ direction. The $V_{xy}^{1\omega}$ exhibits a characteristic parabolic behavior with opposite sign of the quadratic term for the contrary sweeping directions, while $V_{xy}^{2\omega}$ follows a linear dependence with the same sign of slopes, as shown in Figs. 3(c) and 3(d). By substituting the extracted quadratic coefficient and the slope under different $J$ into Eqs. (5)-(6), the $H_{DL}$ are obtained and summarized in Fig. 3(f). The $H_{DL}/J_{SRO}$ is estimated to be 39.65±1.44 Oe/(10$^{11}$ A/m$^2$). Using the similar procedure while the $H$ is applied along $y$ direction, the $H_{FL}$ associated $V_{xy}^{1\omega}$ and $V_{xy}^{2\omega}$ are measured. The $V_{xy}^{2\omega}$ for $H_{FL}$ exhibits the opposite sign of the slopes, as shown in Fig. 3(e), which is consistent with the symmetry of the fieldlike torque. Accordingly, the relationship between $H_{FL}$ and $J$ is plotted in Fig. 3(f), and the $H_{FL}/J$ is estimated to be 2.42±0.34 Oe/(10$^{11}$

A/m$^2$) after deducting the contribution of $H_{Oe}$. The SOT-efficiencies $\xi_{DL}^{PMA}$, $\xi_{FL}^{PMA}$ and spin Hall conductance $\sigma_{SH}^{PMA}$ therefore are confirmed to be 0.210, 0.013 and 4.68×10$^4$ $\hbar$/2e S·m$^{-1}$ in the PMA system with $M_s$ of 499 emu/cm$^3$.

**SOT-induced magnetization switching**

To directly demonstrate practical capability of the SOT that arises from SRO, we further perform current-induced magnetization switching in the SRO/CoPt sample. The measurement configuration is shown in Fig. 4(a). A train of 200 μs pulsed currents $I_{pulse}$ with gradually magnitude in the range between ±24 mA is applied to trigger the magnetization switching. Concurrently, the Hall resistance is acquired under a small DC current $I_{DC}$ of 200 μA that cause negligible influence on the magnetization state. As shown in Fig. 4(b), we observe typical SOT-induced behaviors in the PMA system. Firstly, no switching is detected under a zero bias field due to the mirror symmetry of SOT with respect to normal plane. After applying in-plane bias fields $H_x$ to break the symmetry, deterministic switching occurs at approximately ±19 mA. Further reversing the polarity of $H_x$, the chirality of the loops changes accordingly. These behaviors indicate that the current-induced switching is governed by SOT. The magnitude of the Hall resistance $\Delta R_H$ is partial with that of the AHE curve in Fig. 3(b), likely because the non-uniform pinning hampers a further domain walls displacement in this domain-wall-mediated switching regime. The threshold current density $J_{th}$ of SRO is estimated to be 3.8×10$^{10}$ A/m$^2$, which is about one order of magnitude smaller than those in typical heavy metals [4], such as Pt, Ta, and W. Additionally, to confirm the repeatability and stability of SOT switching using SrRuO$_3$, multiple $I_{pulse}$ with alternative magnitudes of ±22 mA and width of 200 μs are applied to the sample at $H_x$ = ±200 Oe. Fig 4(c) shows the switching results of total 168 pulses, which exhibits highly reproducible SOT-driven response with an almost unchanged $\Delta R_H$, demonstrating the robustness of SRO for current-induced magnetization switching.

**Discussion**

Although previous works have demonstrated that SRO holds novel and highly tunable spin Hall properties, however, there is uncertain magnitude among the reported SOT-efficiencies in IMA system by different measurement techniques. For a reasonable benchmarking, we characterize the SOT properties of high-quality SRO in heterostructures with both IMA and PMA configuration. Since SOT characterization by ST-FMR technique in the previous studies suffers from two main issues of impedance mismatch and phase asynchrony that may introduce undesirable artifacts in the estimated results, we use harmonic Hall voltage technique that can avoid these issues and obtain reasonable results after excluding the extrinsic factors caused from thermal effects and magnetic moment misalignment. As the results listed in table 1, a consistent magnitude of SOT-efficiencies is evaluated to be 0.2 from the both systems, indicating that SRO possesses the pronounced charge-spin conversion efficiency. The subsequently calculated spin Hall conductivity $\sigma_{SH} = \xi_{DL}\sigma_{SRO}$ with the value as high as 4.7×10$^4$ $\hbar$/2e S·m$^{-1}$ in agreement with that of the previous reported magnitude [14]. Moreover, we demonstrate that SOT-driven magnetization switching can be efficiently realized by SRO at room temperature. The threshold current density $J_{th}$ is about 3.8×10$^{10}$ A/m$^2$ in a pulsed time of 200 μs, taking a crucial step for utilizing SRO towards practical SOT applications.

Finally, we compare the SOT properties of SRO with other HMs. As a massively explored system, magnetic alloys combined with HMs, is practically important for realizing SOT device with the advantages of high melting point and non-toxicity. Our study clearly shows that, as listed in table 1, SRO holds comparable $\xi_{DL}$ and $\sigma$, and about one of magnitude smaller $J_{th}$ than that of the typical HMs. Besides, SRO also possesses excellent thermal and chemical stabilities [35], thereby maintains low element diffusivity with the adjacent magnetic alloy layer. Such elemental diffusion should bring adverse impact to SOT efficiency and/or magnetic properties near the heterostructure interface, hindering performance stabilities of practical devices. Therefore, we suggest that by combining these advantages, SRO should be a versatile material for realizing various SOT applications in high quality oxide-metal hybrid system as another choice of HMs, and in all-oxide heterostructures as a standard spin source.

**Conclusion**

We systematically investigate the SOT associated properties of SRO via constructing SRO/magnetic alloy hybrid systems with different magnetization easy axes. One of the main aims of our work is to accurately evaluate the SOT-efficiency in SRO, and make a conclusive comparison with heavy metals. Using harmonic Hall voltage technique in IMA and PMA configuration, we clearly demonstrate that SRO possess a large dampinglike efficiency $\xi_{DL}$ of about 0.2 at room temperature regardless of the magnetic alloys, which is comparable to the typical HMs. The other aim is to demonstrate the current-induced magnetization switching by SRO. We unambiguously observe the SOT-driven magnetization switching behavior, with threshold current density of $3.8\times10^{10}$ A/m$^2$, which is about one magnitude smaller than that of the HMs. From these benchmarks, we suggest that transition metal oxide SRO may bring more diverse choices for SOT applications by utilizing hybrid-oxide/metal and all-oxide systems.

**Methods**

**Film and device fabrication**

Single crystal SRO films of 20 nm are deposited on the (001)-oriented SrTiO$_3$ (STO) substrates by pulsed laser deposition (PLD) with a KrF excimer laser ($\lambda$ = 248 nm). During the deposition, substrate temperature is set at 670 °C and the oxygen partial pressure was 0.1 mbar. The energy density of laser spot is 2.45 J/cm$^2$. After SrRuO$_3$ growth, the films are cooled to room temperature with a rate of 10 °C/min in 1 mbar oxygen atmosphere. Subsequently, a 7-nm-thick Py (Ni$_{81}$Fe$_{19}$) layer is deposited on SrRuO$_3$ under high vacuum, then the IMA system sample of SRO/Py is obtained. For the PMA sample, a CoPt film with a stacking structure of Pt(1nm)/Co(0.5nm)/Pt(0.5nm)/Co(0.5nm)/Pt(1nm) is sputtered on bare SRO by a magnetron sputter at room temperature. The thickness of SRO layer ($t_{SRO}$) is monitored by the *in-situ* Reflection High Energy Electron Diffraction (RHEED) equipped in PLD during the growth processions. And the thicknesses of Py and CoPt layers are estimated by X-Ray Reflectivity (XRR) scan.

**Device fabrication and measurements**

For electrical measurements, the samples are fabricated into Hall bar devices using standard photolithography and argon ion etching techniques. The Hall bar pattern has channel dimension of 10μm in current-interface width and 5μm in voltage-interface width. After that, 10-nm Ti and 50-nm Au are successively deposited on the Hall bar pattern by electron-beam evaporation. The saturation magnetization $M_S$ of magnetic layers is measured by superconducting quantum interference device (SQUID, Quantum Design). The electrical measurements are performed on the home-build low-temperature and high magnetic field electronic test system. The fixed AC current with frequency 133.73 Hz and the pulse DC current are applied by Keithley 6221 current source, and the constant DC current is generated by Keithley 6220 current source. The harmonic voltage signals and DC voltage signal are respectively measured by SR830 DSP Lock-in amplifiers and Keithley 2182A nanovoltmeter.

**Acknowledgments**

This work was supported by the National Key Research and Development Program of China (Nos. 2017YFA0303600, 2019YFA0307800), the National Natural Science Foundation of China (Nos. 12174406, 11874367, 51931011), the Key Research Program of Frontier Sciences, Chinese Academy of Sciences (No. ZDBS-LY-SLH008), K.C.Wong Education Foundation (GJTD-2020-11), the 3315 Program of Ningbo, the Natural Science Foundation of Zhejiang province of China (No. LR20A040001), the Beijing National Laboratory for Condensed Matter Physics.

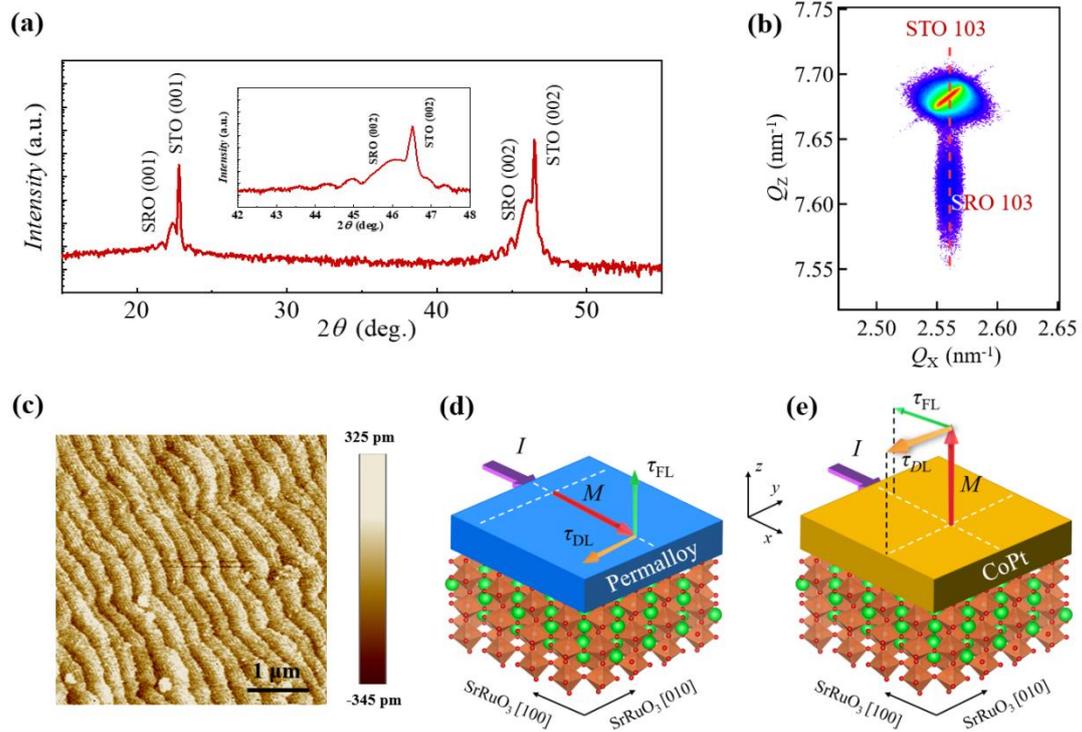

**Fig. 1 SrRuO₃-based heterostructures. a** XRD ω-2θ scan for the SrRuO₃ film grown on the SrTiO₃ (001) substrate. Inset shows the zoom-in range around (002) peaks of SrRuO₃ and SrTiO₃. **b** Reciprocal space mapping around (103) peak for the prepared film. **c** An AFM image of the film. **d-e** Schematic diagram of spin orbit torques in SrRuO₃-based systems integrated with **(d)** permalloy and **(e)** CoPt, which show in-plane and perpendicular magnetic anisotropy, respectively. The purple, red, orange and green arrows represent charge current (*I*), magnetization (*M*), dampinglike and filedlike torques ($\tau_{DL}$ and $\tau_{FL}$), respectively. The *x*, *y* and *z* axes in Cartesian coordinates are parallel to the [100], [010] and [001] crystal directions of the SrRuO₃, respectively.

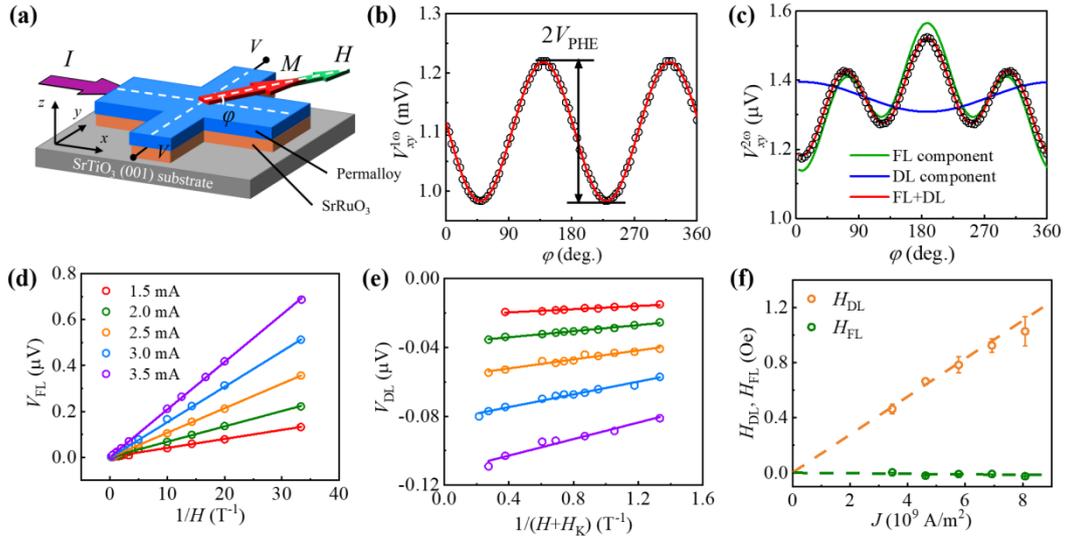

**Fig. 2 Harmonic Hall voltage measurement of the SrRuO$_3$/Py sample. a** A schematic diagram for the measurement. The current $I$ flows along the $x$-axis. $\varphi$ is defined as the angle between $I$ and the external magnetic field $H$. **b-c** Typical first (**b**) and second (**c**) harmonic Hall voltages ($V_{xy}^{1\omega}$ and $V_{xy}^{2\omega}$) measured at $H = 1000$ Oe and $I = 3.5$ mA. The $V_{xy}^{2\omega}$ consists of the dampinglike and fieldlike components ($V_{FL}$ and $V_{DL}$). **d** Linear fitting of $V_{FL}$ against $1/H$ measured at different $I$. **e** Linear fitting of $V_{DL}$ against $1/(H+H_K)$ measured at different $I$. **f** The equivalent fields, i.e. $H_{DL}$ and $H_{FL}$, as a function of current density $J$ and corresponding linearly fitted lines.

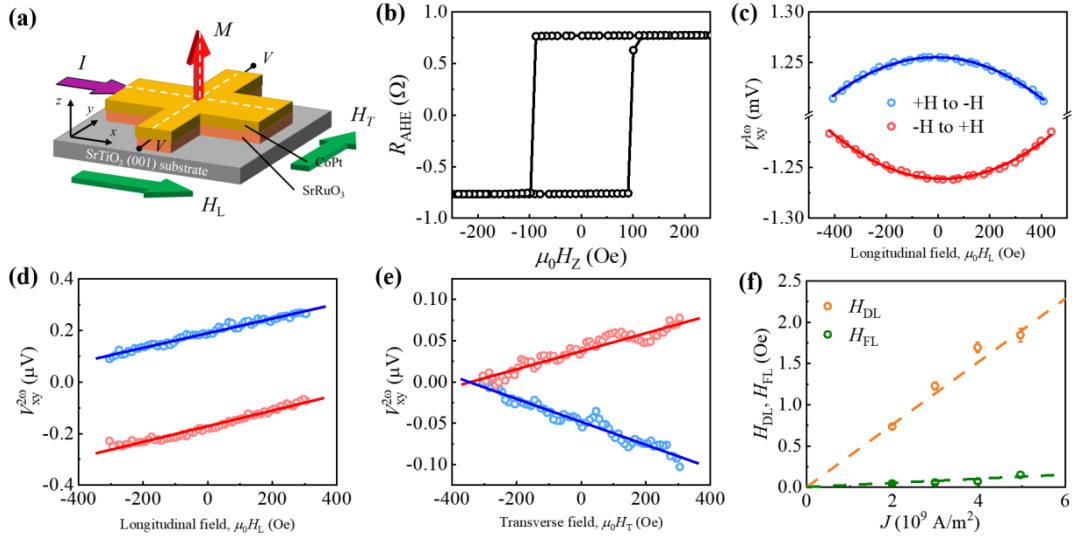

**Fig. 3 Harmonic Hall voltage measurement of the SrRuO$_3$/CoPt sample. a** A schematic diagram for the measurement. **b** The measured anomalous Hall resistance $R_{AHE}$ as a function of out-of-plane magnetic field $H_Z$. **c-d** The first **(c)** and second **(d)** harmonic Hall voltages dependent on in-plane longitudinal field $H_L$ measured at $I = 3$mA. When the first harmonic Hall voltage is measured. **e** The second harmonic Hall voltage dependent on the in-plane transverse field $H_T$ measured at $I = 3$mA. **f** The equivalent fields, i.e. $H_{DL}$ and $H_{FL}$, as a function of current density $J$ and corresponding linearly fitted lines.

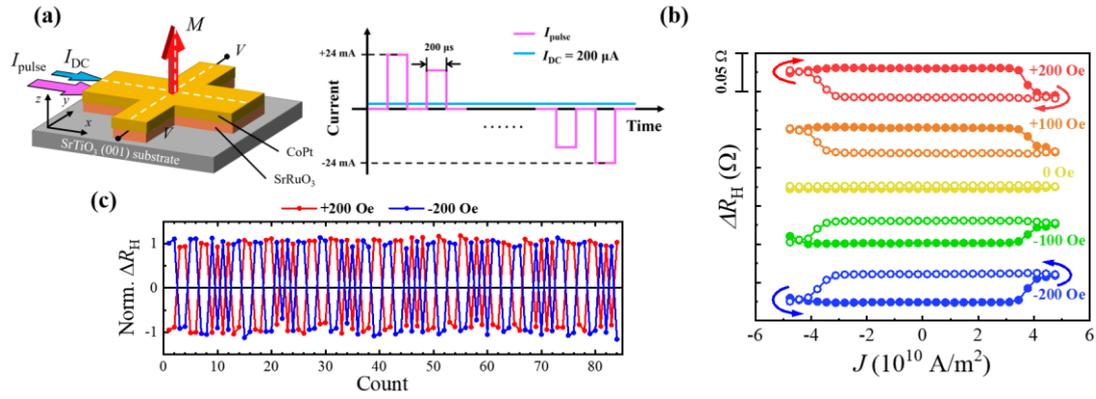

**Fig. 4 Spin-orbit torque induced perpendicular magnetization switching in SrRuO$_3$/CoPt hybrid structure. a** Left panel: the schematic diagram of the set-up for the switching. The pulsed current $I_{\text{pulse}}$ and the DC constant current $I_{\text{DC}}$ are used to switch and read the magnetization, respectively. Right panel: the sequence diagram of $I_{\text{pulse}}$ (+24mA ~ -24mA) and $I_{\text{DC}}$ (200 μA). **b** Magnetization switching driven by $I_{\text{pulse}}$ under different external magnetic field $H_x$. $\Delta R_H$ represents the change of Hall resistance. **c** SOT-induced magnetization reversal with a switching current $I_{\text{pulse}}$ (± 22mA, 200 μs) and an assistant field $H_x$ (±200Oe).

| Sample | Magnetic Anisotropy | Technique | $\xi_{DL}$ | $\xi_{FL}$ | $\rho_{SOC}$ (μΩ·cm) | $\sigma_{SH}$ ($\times 10^4\ \hbar/2e$ S m$^{-1}$) | $J_{th}$ ($\times 10^{10}$ A/m$^2$) | Reference |
|---|---|---|---|---|---|---|---|---|
| SRO(20)/Py(7) | IMA | Harmonic | 0.18 | -0.003 | 156 | 11 | | this work |
| SRO(20)/CoPt(3.5) | PMA | Harmonic | 0.21 | 0.013 | 450 | 4.7 | 3.8 | this work |
| SRO(20)/Py(4) on STO(001) | IMA | ST-FMR | 0.14 | | 120 | 23 | | [15] |
| SRO(20)/Py(4) on STO(001) | IMA | Harmonic | 0.035 | | 120 | 5.8 | | [15] |
| SRO(6)/Py(6) on STO(001) | IMA | ST-FMR | 0.49 | | 810 | 5.7 | | [16] |
| Pt(6)/Py(4) | IMA | ST-FMR | 0.056 | | 20 | 28 | | [47] |
| Pt(3)/Co(0.6) | PMA | Harmonic | 0.16 | | 36 | 44 | 82 | [48] |
| Pt(4)/Co(0.85) | PMA | Harmonic | 0.2 | -0.05 | 50 | 40 | 32 | [49] |
| W(6)/CoFeB(5) (β-phase) | IMA | ST-FMR | 0.3 | | 170 | 17 | 18 | [50] |
| Ta(8)/CoFeB(5) (β-phase) | PMA | ST-FMR | 0.15 | | 190 | 7.9 | 48 | [52] |
| Ta(13)/Py(4) (β-phase) | IMA | ST-FMR | -0.37 | 0.17 | 122 | -6.6 | | [51] |
| Ta(13)/Py(4) (Mixed-phase) | IMA | ST-FMR | -0.52 | -0.06 | 45 | -50 | | [51] |

**Table 1** Comparison of the SOT-efficiencies and spin Hall conductivity in different systems. $\xi_{DL(FL)}$ is the dampinglike (fieldlike) SOT-efficiency, $\rho_{SOC}$ is the resistivity of SOC layer, $\sigma_{SH}$ and $J_{th}$ present the spin Hall conductivity and threshold current density, respectively.